\documentclass[a4paper]{article}

\usepackage{INTERSPEECH2020}

\title{Predicting Different Acoustic Features from EEG and towards direct synthesis of Audio Waveform from EEG}
\name{Gautam Krishna, Co Tran, Mason Carnahan, Ahmed H Tewfik}
\address{
  Brain Machine Interface Lab, The University of Texas at Austin
  }
\email{}

\begin{document}

\maketitle
\begin{abstract}
In \cite{krishna2020synthesis,krishna2020advancing} authors provided preliminary results for synthesizing speech from electroencephalography (EEG) features where they first predict acoustic features from EEG features and then the speech is reconstructed from the predicted acoustic features using griffin lim reconstruction algorithm. In this paper we first introduce a deep learning model that takes raw EEG waveform signals as input and directly produces audio waveform as output. We then demonstrate predicting 16 different acoustic features from EEG features. We demonstrate our results for both spoken and listen condition in this paper. The results presented in this paper shows how different acoustic features are related to non-invasive neural EEG signals recorded during speech perception and production. 

\end{abstract}
\noindent\textbf{Index Terms}: electroencephalography (EEG), speech synthesis, deep learning, technology accessibility 

\section{Introduction}

Synthesizing speech from neural signals might help people with speaking disabilities or disorders like aphasia, stuttering etc to use virtual personal assistants like Bixby, Siri, Alexa etc there by helping in improving technology accessibility. It might also help people who can't produce any voice at all like for example people recovering from sever stroke or amyotrophic lateral sclerosis (ALS) with speech restoration. So far most promising results on synthesizing speech from neural signals were obtained using speech synthesis from invasive electrocorticography (ECoG) neural signals where a subject need to undergo a brain surgery to implant ECoG electrodes on the surface of the brain. The ECoG signals offer very high temporal resolution, good spatial resolution and signal to noise ratio.
The works described in references  \cite{anumanchipalli2019speech,herff2019generating,angrick2019speech} provides results on synthesizing intelligible speech from ECoG recordings. One major limitation of ECoG  approach is the requirement of brain surgery to implant the electrodes which makes ECoG based speech prosthetic systems difficult to deploy and study. 

On the other hand we have electroencephalography (EEG) signals which is a non-invasive way of measuring electrical activity of human brain where EEG sensors are placed on the scalp of the subject to obtain the EEG readings. The EEG signals offers high temporal resolution like ECoG signals, however the major disadvantage of EEG is the spatial resolution and signal to noise ratio offered are poor compared to ECoG. The non-invasive nature of EEG makes it easy to study and deploy compared to ECoG. Due to the poor signal to noise ratio offered by EEG signals the performance or accuracy of EEG based brain computer interface (BCI) systems are poor compared to ECoG based BCI systems. In  \cite{krishna20,krishna2019speech,krishna2019state} authors demonstrated the feasibility of performing continuous and silent speech recognition using EEG signals in presence and absence of background noise. They used principles of automatic speech recognition (ASR) to translate the EEG recordings to text. Their results were demonstrated for limited English vocabulary. Similarly in \cite{ramsey2017decoding,martin2016word} authors demonstrated speech recognition using ECoG signals with better accuracy. 
One approach to synthesize speech from EEG signals would be first use an EEG based continuous speech recognizer to produce text from EEG and then use an existing state of art text to speech (TTS) system to convert text to sound. However this approach suffers from following limitations - First of all the word error rates reported by current existing EEG based continuous speech recognizer is very high during test time \cite{krishna20,krishna2019state,krishna2019improving} and then using a TTS system at the final stage in the pipeline will introduce additional latency. It is possible to produce personalized sound using speaker dependent TTS but it is not clear how these systems will perform if the speech is broken or distorted like in the case of speech from an aphasia or stuttering patient. 
So an alternative approach would be to produce speech directly from EEG signals.  In \cite{krishna2020synthesis} authors provided preliminary results for synthesizing speech from EEG features where they reconstruct speech waveform from the predicted acoustic features, however they were not able to produce intelligible speech. In \cite{krishna2020advancing} authors improved the results presented in \cite{krishna2020synthesis} by using attention mechanism \cite{luong2015effective} between the encoder and decoder in their model and they also identified the right sampling frequency to extract features to produce more intelligible speech compared to \cite{krishna2020synthesis}. However in \cite{krishna2020advancing} the speech waveform constructed from the predicted acoustic features were noisy.

In this paper we introduce a deep learning model that takes raw EEG waveform signals as input and directly outputs speech or audio waveform. We show that by this approach it is possible to produce audio waveform from EEG signal with less noise compared to \cite{krishna2020advancing}. We were not able to produce intelligible audio but the results presented in this paper along with results from \cite{krishna2020advancing} shows further evidence on the future possibility of producing high quality intelligible audio from EEG. We finally demonstrate predicting sixteen different acoustic features from EEG features in this paper. In \cite{krishna2020synthesis,krishna2020advancing} authors demonstrated predicting only one type of acoustic feature namely mel frequency cepstral coefficients (MFCCs) from EEG features.

\section{Deep Learning Models}
Figure 1 explains the architecture of our speech synthesis model. The model consists of a temporal convolutional network (TCN) \cite{bai2018empirical} layer with 256 filters which takes raw EEG waveform signals from 31 channels as input and the features extracted by the TCN layer is up sampled at a rate of 15 (5 times 3). A dropout regularization \cite{srivastava2014dropout} with dropout rate 0.2 and a TCN layer with 32 filters is applied before the final up sampling layer. The up sampled features are then passed to a time distributed dense layer consisting of linear activation function with one hidden unit which directly outputs the audio waveform.
The model was trained for 5000 epochs using adam \cite{kingma2014adam} optimizer with a batch size of 100. We used mean squared error (MSE) as the loss function for the model. 

For predicting different acoustic features from EEG features we used a gated recurrent unit (GRU) \cite{chung2014empirical} based regression model. The model consists of a single layer of GRU with 128 hidden units which takes EEG features of dimension 30 as input and the GRU output features are passed to a dropout regularization with dropout rate 0.2 and then to a time distributed dense layer with linear activation function. The number of hidden units in the time distributed dense layer was same as the dimension of the acoustic feature type the model was predicting for the given input EEG features. The model was trained for 500 epochs with a batch size of 100. The MSE was used as the loss function for the model with adam as the optimizer.

\begin{figure}[h]
\begin{center}
\includegraphics[height=6.5cm, width=\linewidth,trim={0.1cm 0.1cm 0.1cm 0.1cm}]{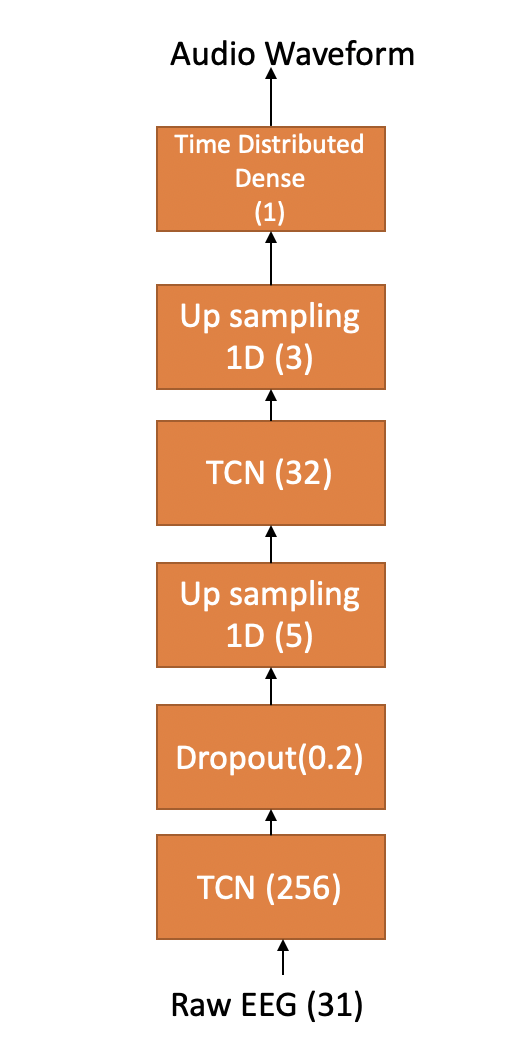}
\caption{Speech Synthesis Model} 
\label{1vsall}
\end{center}
\end{figure}


\section{Data Sets used for performing experiments}
We used the Data set used by authors in \cite{krishna2020synthesis} for this work. In their experiment they first ask four subjects to listen to natural English utterances and speak out loud the utterances that they listened to. The EEG signals were recorded in parallel while they were listening to the utterances known as listen EEG and also the EEG signals were recorded in parallel while they were speaking out the utterances known as spoken EEG.  The listening utterances and their speech was also recorded. In this paper by the term listen condition we refer to the problem of predicting listening utterance waveform or listening utterance different acoustic features from listen EEG waveform or listen EEG features and by the term spoken condition we refer to the problem of predicting
spoken speech waveform or spoken speech different acoustic features from spoken EEG waveform or spoken EEG features. 
More details of the experiment design for collecting simultaneous speech and EEG data are covered in \cite{krishna2020synthesis}.  

They used Brain product's ActiChamp EEG amplifier. Their EEG cap had 32 wet EEG electrodes including one electrode as ground. It is based on standard 10-20 EEG sensor placement method for 32 electrodes.

For each experiment set we used 80\% of the data as training set, remaining 10\% as validation set and rest 10\% as test set. The train-test split was done randomly. There was no overlap between training, testing and validation set. The way we splitted data in this work is exactly same as the method used by authors in \cite{krishna2020synthesis,krishna2020advancing}.  

\section{EEG and Speech feature extraction details}
We followed the same EEG preprocessing methods used by authors in \cite{krishna2020advancing}. 
The EEG signals were sampled at 1000Hz and a fourth order IIR band pass filter with cut off frequencies 0.1Hz and 70Hz was applied. A notch filter with cut off frequency 60 Hz was used to remove the power line noise.
The EEGlab's \cite{delorme2004eeglab} Independent component analysis (ICA) toolbox was used to remove other biological signal artifacts like electrocardiography (ECG), electromyography (EMG), electrooculography (EOG) etc from the EEG signals. These EEG signals after artifact removal or correction were fed to the speech synthesis model described in Figure 1. Thus the model in Figure 1 took EEG waveform from 31 channels as input as the 32nd channel was ground.   

We then extracted five statistical features for EEG, namely root mean square, zero crossing rate, moving window average, kurtosis and power spectral entropy \cite{zhang2008feature} as explained by authors in \cite{krishna2019speech,krishna20} for the problem of predicting different acoustic features from EEG features. 

The recorded speech signal was sampled at 16KHz frequency. It was then down sampled to 15KHz frequency. We then extracted 16 different acoustic features namely power spectrogram of dimension 12, Constant-Q chromagram of dimension 12, chroma based features \cite{muller2011chroma} of dimension 12, mel-scaled spectrogram of dimension 128, root-mean-square (RMS) of dimension one, spectral centroid of dimension one, spectral bandwidth of dimension one, spectral contrast of dimension 7, 
spectral flatness of dimension one, roll-off frequency of dimension one, coefficients of fitting a first order polynomial to the columns of a spectrogram of dimension 2, tonal centroid features \cite{harte2006detecting} of dimension 6,  zero-crossing rate of dimension one, tempogram of dimension 384, loudness of dimension one and pitch of dimension one. 

The EEG features and all the audio features were extracted at a sampling frequency of 31 Hz. In \cite{krishna2020advancing,krishna2020synthesis} authors extracted features at 100Hz, 32 Hz.

\begin{figure}[h]
\begin{center}
\includegraphics[height=6.0cm, width=\linewidth,trim={0.1cm 0.1cm 0.1cm 0.1cm}]{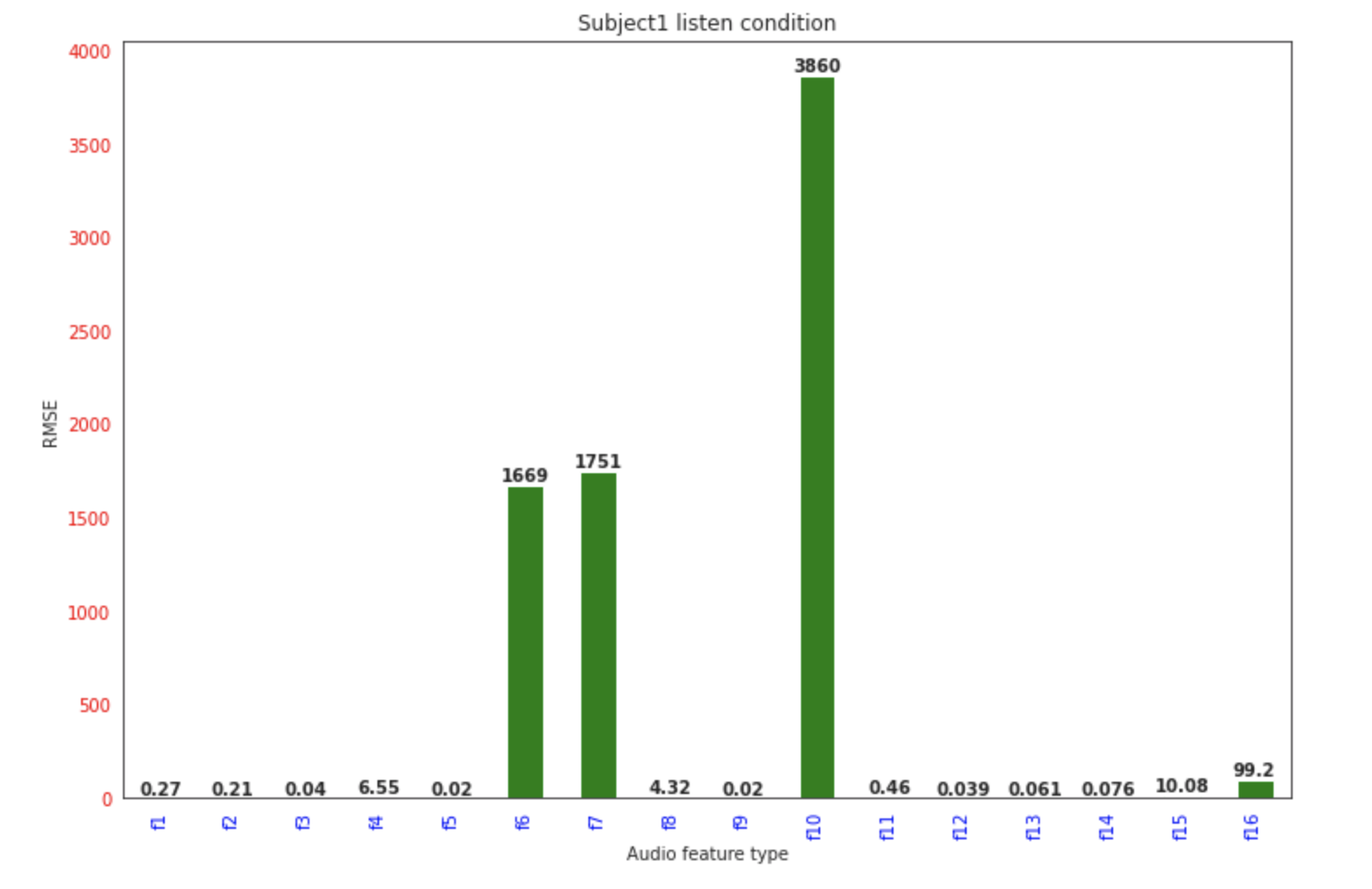}
\caption{Test time results for predicting different types of acoustic features from EEG features for subject 1 for \textbf{Listen condition}} 
\label{1vsall}
\end{center}
\end{figure}

\begin{figure}[h]
\begin{center}
\includegraphics[height=6.0cm, width=\linewidth,trim={0.1cm 0.1cm 0.1cm 0.1cm}]{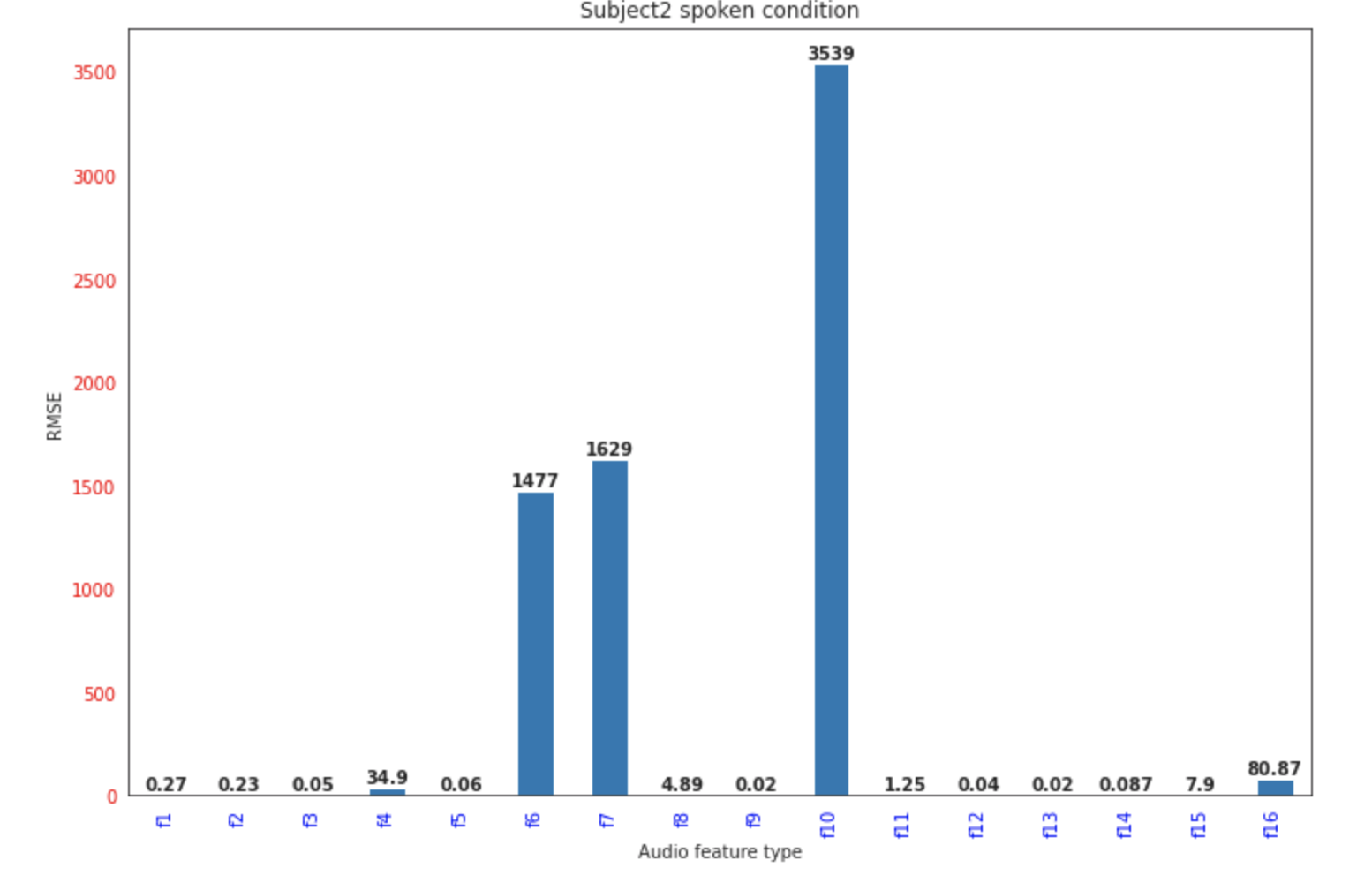}
\caption{Test time results for predicting different types of acoustic features from EEG features for subject 2 for \textbf{Spoken condition}} 
\label{1vsall}
\end{center}
\end{figure}

\begin{figure}[h]
\begin{center}
\includegraphics[height=6.5cm, width=\linewidth,trim={0.1cm 0.1cm 0.1cm 0.1cm}]{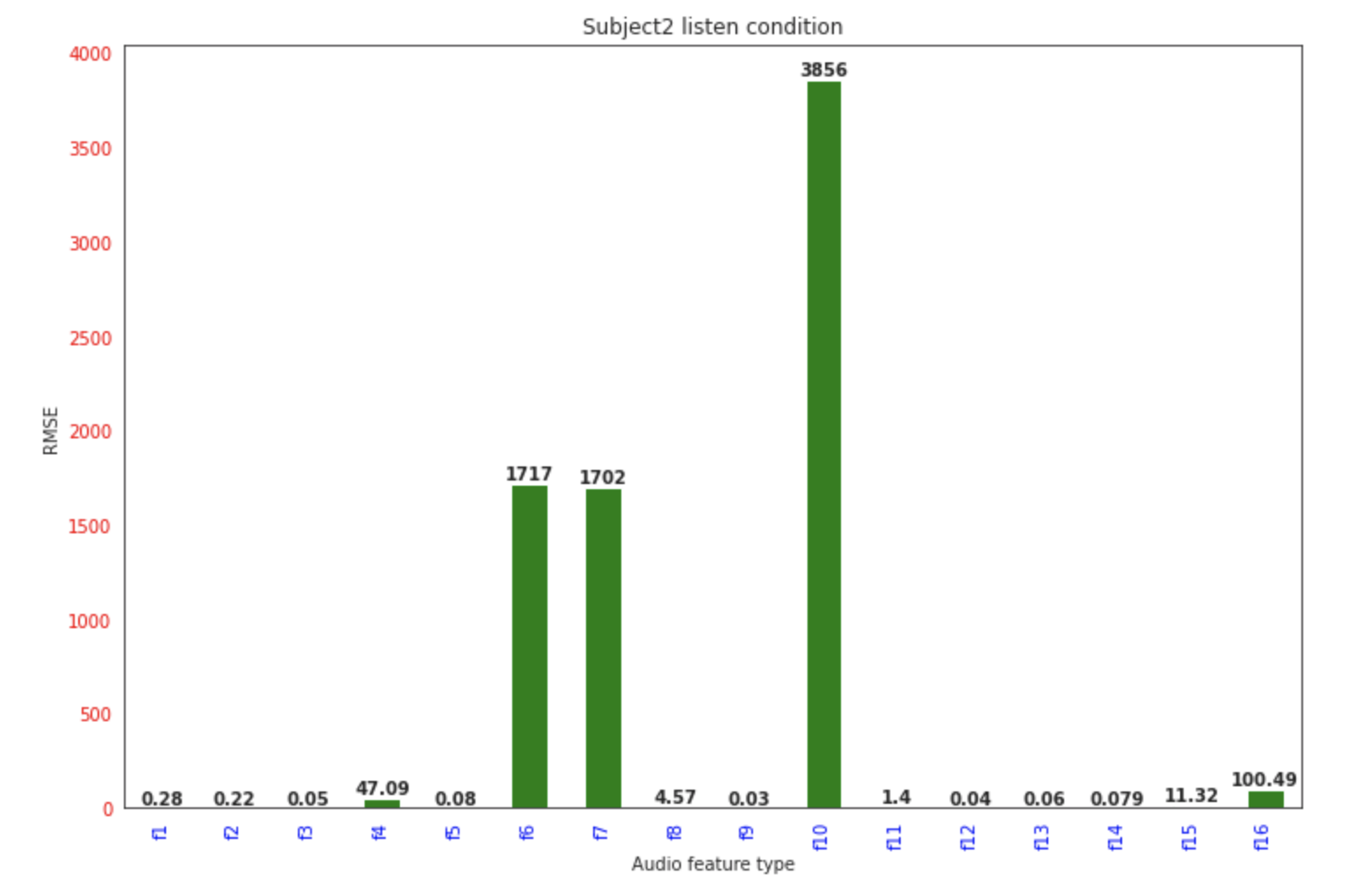}
\caption{Test time results for predicting different types of acoustic features from EEG features for subject 2 for \textbf{Listen condition}} 
\label{1vsall}
\end{center}
\end{figure}

\section{EEG Feature Dimension Reduction Algorithm Details}

We used kernel principal component analysis (KPCA) \cite{mika1999kernel} to de-noise the EEG feature space by performing dimension reduction for the extracted 155 EEG features (31 channel X 5 features) as explained by authors in \cite{krishna2019state,krishna2020synthesis,krishna2020advancing}. 
By following the dimension reduction method explained by authors in \cite{krishna2019state,krishna20} we reduced 155 EEG features to a dimension of 30. More details of explained variance plot used to identify the right feature dimension is covered in \cite{krishna2019state,krishna20}. We used a polynomial kernel with degree 3 for performing kernel PCA.

\section{Results}

In \cite{krishna2020advancing} authors used mel cepstral distortion (MCD) \cite{kominek2008synthesizer} as the performance metric during test time but since here we are directly producing audio waveform as the output we used root mean squared error (RMSE) between the predicted audio waveform and ground truth audio waveform from test set as performance metric during test time for speech synthesis problem. The Figure 6 shows the obtained test time results for speech synthesis for four subjects for both spoken and listen condition. For spoken condition we observed lowest RMSE value of 0.583 for subject 2 and for listen condition we observed lowest RMSE value of 0.489 again for subject 2.  The Figure 7 shows the visualization of the predicted waveform and ground truth waveform from test set for a sample for subject 1 for spoken condition. As seen from the plots, our predicted waveform was comparable with the waveform reconstructed from the predicted MFCC or acoustic features explained in \cite{krishna2020advancing} in terms of capturing very broad characteristics of the actual waveform for the same transcript from the same subject, however we observed that our predicted waveform had amplitude values amplified by a bigger margin compared to the predicted waveform explained in \cite{krishna2020advancing}. Figures 8 and 9 shows the corresponding spectrogram for the actual and predicted waveform.
As seen from the spectrograms, the predicted and actual waveform shared some features in common in frequency domain.

\begin{figure}[h]
\begin{center}
\includegraphics[height=5.5cm, width=\linewidth,trim={0.1cm 0.1cm 0.1cm 0.1cm}]{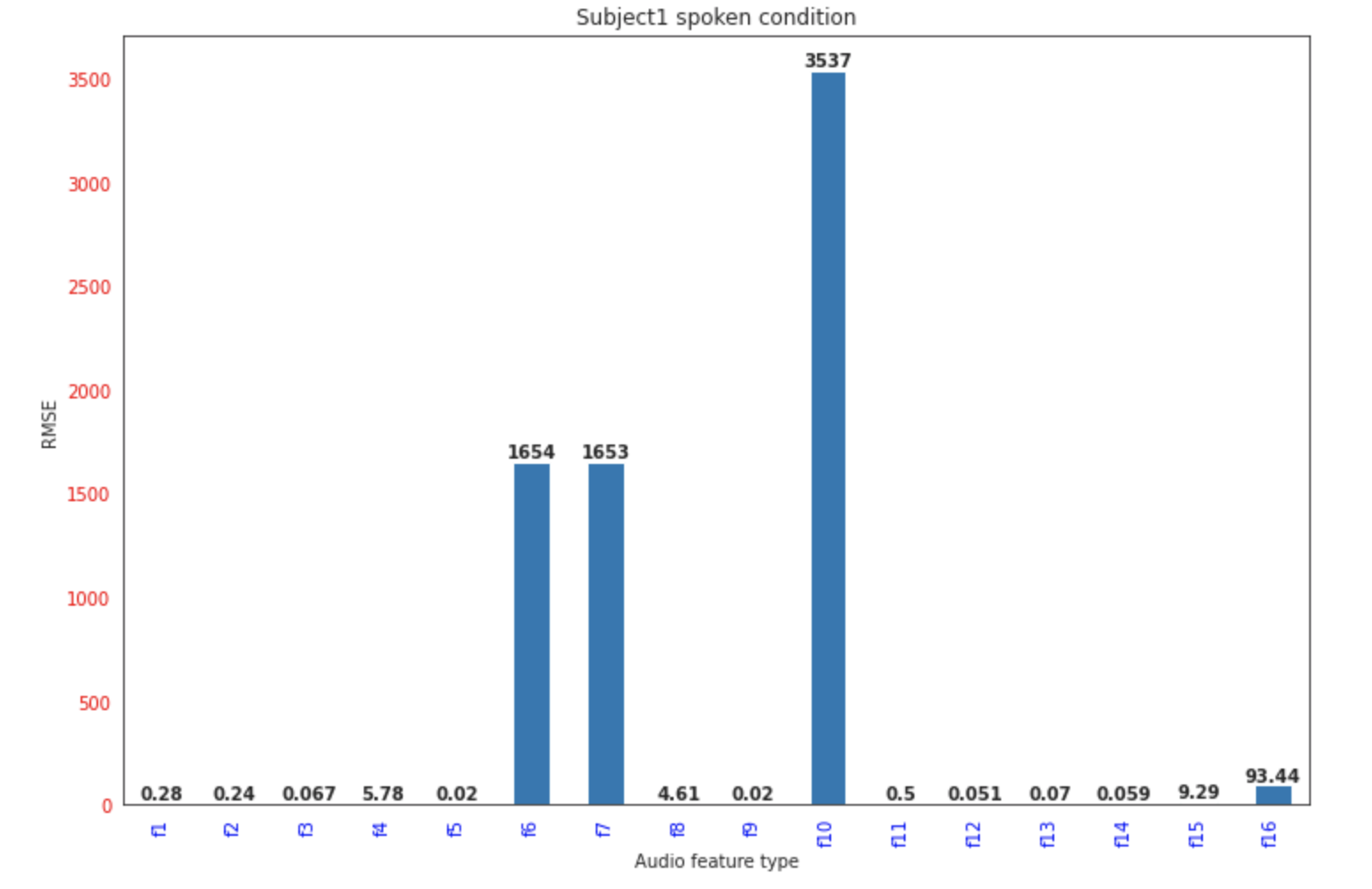}
\caption{Test time results for predicting different types of acoustic features from EEG features for subject 1 for \textbf{Spoken condition}} 
\label{1vsall}
\end{center}
\end{figure}

For predicting different acoustic features from EEG features we again used RMSE as the performance metric during test time. Figures 5 and 2 shows the test time results for subject 1 for spoken and listen condition. Figures 3 and 4 shows the test time results for subject 2 for spoken and listen condition. In the plots x axis label f1 corresponds to first type of acoustic feature (ie : power spectrogram of dimension 12), f2 corresponds to second type of acoustic feature (ie: Constant-Q chromagram of dimension 12) and so on. As seen from the plots we observed highest RMSE value for predicting acoustic feature type 10 or f10 (ie: roll-off frequency of dimension one) for subject 1 and 2 for both listen and spoken condition. Similar results were obtained for remaining subjects 3 and 4 as well.

\begin{figure}[h]
\begin{center}
\includegraphics[height=6.5cm, width=\linewidth,trim={0.1cm 0.1cm 0.1cm 0.1cm}]{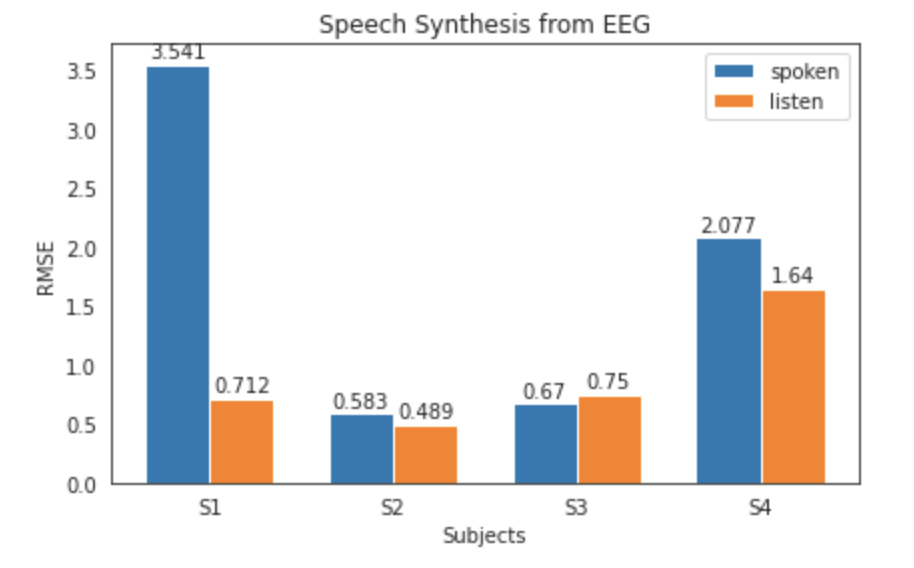}
\caption{Speech synthesis test time results for four subjects} 
\label{1vsall}
\end{center}
\end{figure}

\begin{figure}[h]
\begin{center}
\includegraphics[height=6.5cm, width=\linewidth,trim={0.1cm 0.1cm 0.1cm 0.1cm}]{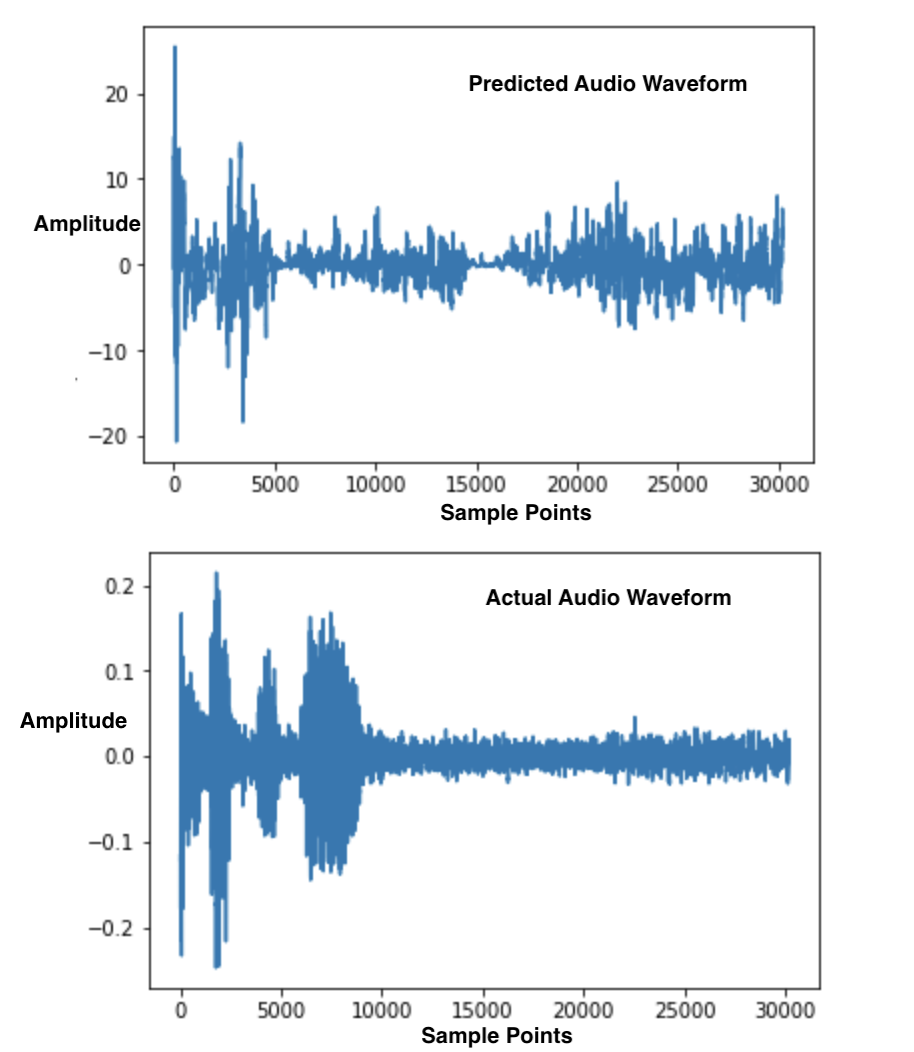}
\caption{Speech synthesis test time result for subject 1. The text corresponding to actual waveform was 'Hi Bixby'} 
\label{1vsall}
\end{center}
\end{figure}

\begin{figure}[h]
\begin{center}
\includegraphics[height=5.0cm, width=\linewidth,trim={0.1cm 0.1cm 0.1cm 0.1cm}]{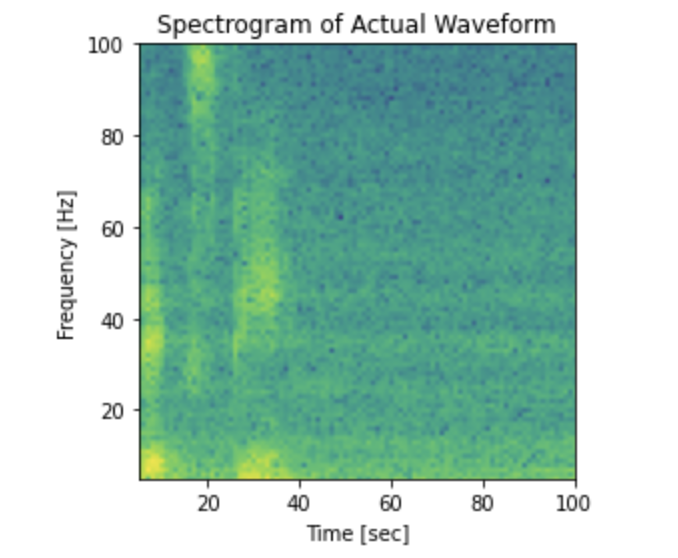}
\caption{Spectrogram of Actual Waveform. The text corresponding to actual waveform was 'Hi Bixby'} 
\label{1vsall}
\end{center}
\end{figure}

\begin{figure}[h]
\begin{center}
\includegraphics[height=5.0cm, width=\linewidth,trim={0.1cm 0.1cm 0.1cm 0.1cm}]{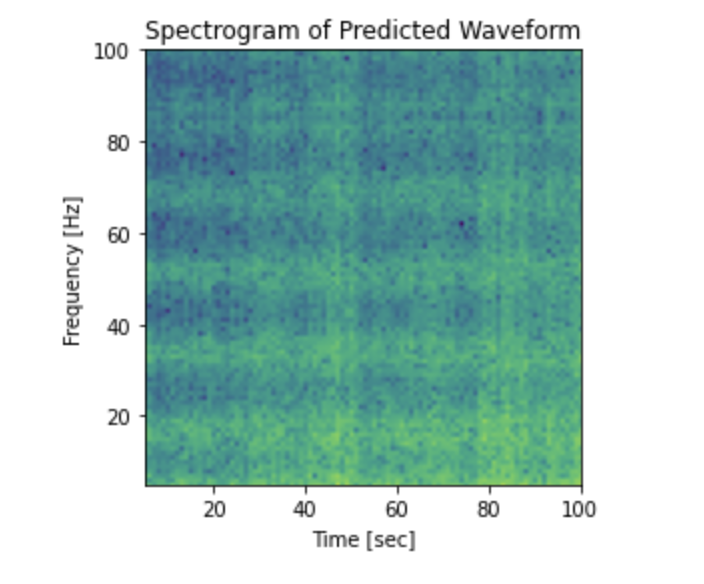}
\caption{Spectrogram of corresponding predicted Waveform} 
\label{1vsall}
\end{center}
\end{figure}

\section{Conclusion and Future work}

In this paper we demonstrated generating audio waveform directly from EEG waveform with low RMSE values during test time using deep learning model. We further demonstrated predicting 16 different types of acoustic features from EEG features. Even though we were able to generate audio waveform only with very broad characteristics of the actual waveform, our generated waveform was comparable to the waveform explained in \cite{krishna2020advancing} by visual inspection in terms of capturing the broad characteristics of the actual audio. The results presented in this paper along with the results explained in \cite{krishna2020advancing} further advances the research on speech synthesis using EEG. 

Our future work will focus on developing highly intelligible audio waveform from EEG signals by overcoming the limitations of this paper and \cite{krishna2020advancing}.

\section{Acknowledgement} 
We would like to thank Kerry Loader and Rezwanul Kabir from Dell, Austin, TX for donating us the GPU to train the models used in this work.

\bibliographystyle{IEEEtran}

\bibliography{mybib}

\begin{thebibliography}{10}
\providecommand{\url}[1]{#1}
\csname url@samestyle\endcsname
\providecommand{\newblock}{\relax}
\providecommand{\bibinfo}[2]{#2}
\providecommand{\BIBentrySTDinterwordspacing}{\spaceskip=0pt\relax}
\providecommand{\BIBentryALTinterwordstretchfactor}{4}
\providecommand{\BIBentryALTinterwordspacing}{\spaceskip=\fontdimen2\font plus
\BIBentryALTinterwordstretchfactor\fontdimen3\font minus
  \fontdimen4\font\relax}
\providecommand{\BIBforeignlanguage}[2]{{%
\expandafter\ifx\csname l@#1\endcsname\relax
\typeout{** WARNING: IEEEtran.bst: No hyphenation pattern has been}%
\typeout{** loaded for the language `#1'. Using the pattern for}%
\typeout{** the default language instead.}%
\else
\language=\csname l@#1\endcsname
\fi
#2}}
\providecommand{\BIBdecl}{\relax}
\BIBdecl

\bibitem{krishna2020synthesis}
G.~Krishna, C.~Tran, Y.~Han, M.~Carnahan, and A.~Tewfik, ``Speech synthesis
  using eeg,'' in \emph{Acoustics, Speech and Signal Processing (ICASSP), 2020
  IEEE International Conference on}.\hskip 1em plus 0.5em minus 0.4em\relax
  IEEE, 2020.

\bibitem{krishna2020advancing}
G.~Krishna, C.~Tran, M.~Carnahan, and A.~Tewfik, ``Advancing speech synthesis
  using eeg,'' \emph{arXiv preprint arXiv:2004.04731}, 2020.

\bibitem{anumanchipalli2019speech}
G.~K. Anumanchipalli, J.~Chartier, and E.~F. Chang, ``Speech synthesis from
  neural decoding of spoken sentences,'' \emph{Nature}, vol. 568, no. 7753, p.
  493, 2019.

\bibitem{herff2019generating}
C.~Herff, L.~Diener, M.~Angrick, E.~M. Mugler, M.~C. Tate, M.~Goldrick,
  D.~Krusienski, M.~W. Slutzky, and T.~Schultz, ``Generating natural,
  intelligible speech from brain activity in motor, premotor and inferior
  frontal cortices,'' \emph{Frontiers in Neuroscience}, vol.~13, p. 1267, 2019.

\bibitem{angrick2019speech}
M.~Angrick, C.~Herff, E.~Mugler, M.~C. Tate, M.~W. Slutzky, D.~J. Krusienski,
  and T.~Schultz, ``Speech synthesis from ecog using densely connected 3d
  convolutional neural networks,'' \emph{Journal of neural engineering},
  vol.~16, no.~3, p. 036019, 2019.

\bibitem{krishna20}
G.~Krishna, C.~Tran, M.~Carnahan, and A.~Tewfik, ``Advancing speech recognition
  with no speech or with noisy speech,'' in \emph{2019 27th European Signal
  Processing Conference (EUSIPCO)}.\hskip 1em plus 0.5em minus 0.4em\relax
  IEEE, 2019.

\bibitem{krishna2019speech}
G.~Krishna, C.~Tran, J.~Yu, and A.~Tewfik, ``Speech recognition with no speech
  or with noisy speech,'' in \emph{Acoustics, Speech and Signal Processing
  (ICASSP), 2019 IEEE International Conference on}.\hskip 1em plus 0.5em minus
  0.4em\relax IEEE, 2019.

\bibitem{krishna2019state}
G.~Krishna, Y.~Han, C.~Tran, M.~Carnahan, and A.~H. Tewfik, ``State-of-the-art
  speech recognition using eeg and towards decoding of speech spectrum from
  eeg,'' \emph{arXiv preprint arXiv:1908.05743}, 2019.

\bibitem{ramsey2017decoding}
N.~Ramsey, E.~Salari, E.~Aarnoutse, M.~Vansteensel, M.~Bleichner, and
  Z.~Freudenburg, ``Decoding spoken phonemes from sensorimotor cortex with
  high-density ecog grids,'' \emph{Neuroimage}, 2017.

\bibitem{martin2016word}
S.~Martin, P.~Brunner, I.~Iturrate, J.~d.~R. Mill{\'a}n, G.~Schalk, R.~T.
  Knight, and B.~N. Pasley, ``Word pair classification during imagined speech
  using direct brain recordings,'' \emph{Scientific reports}, vol.~6, p. 25803,
  2016.

\bibitem{krishna2019improving}
G.~Krishna, C.~Tran, M.~Carnahan, Y.~Han, and A.~H. Tewfik, ``Improving eeg
  based continuous speech recognition,'' \emph{arXiv preprint
  arXiv:1911.11610}, 2019.

\bibitem{luong2015effective}
M.-T. Luong, H.~Pham, and C.~D. Manning, ``Effective approaches to
  attention-based neural machine translation,'' \emph{arXiv preprint
  arXiv:1508.04025}, 2015.

\bibitem{bai2018empirical}
S.~Bai, J.~Z. Kolter, and V.~Koltun, ``An empirical evaluation of generic
  convolutional and recurrent networks for sequence modeling,'' \emph{arXiv
  preprint arXiv:1803.01271}, 2018.

\bibitem{srivastava2014dropout}
N.~Srivastava, G.~Hinton, A.~Krizhevsky, I.~Sutskever, and R.~Salakhutdinov,
  ``Dropout: a simple way to prevent neural networks from overfitting,''
  \emph{The journal of machine learning research}, vol.~15, no.~1, pp.
  1929--1958, 2014.

\bibitem{kingma2014adam}
D.~P. Kingma and J.~Ba, ``Adam: A method for stochastic optimization,''
  \emph{arXiv preprint arXiv:1412.6980}, 2014.

\bibitem{chung2014empirical}
J.~Chung, C.~Gulcehre, K.~Cho, and Y.~Bengio, ``Empirical evaluation of gated
  recurrent neural networks on sequence modeling,'' \emph{arXiv preprint
  arXiv:1412.3555}, 2014.

\bibitem{delorme2004eeglab}
A.~Delorme and S.~Makeig, ``Eeglab: an open source toolbox for analysis of
  single-trial eeg dynamics including independent component analysis,''
  \emph{Journal of neuroscience methods}, vol. 134, no.~1, pp. 9--21, 2004.

\bibitem{zhang2008feature}
A.~Zhang, B.~Yang, and L.~Huang, ``Feature extraction of eeg signals using
  power spectral entropy,'' in \emph{2008 International Conference on
  BioMedical Engineering and Informatics}.\hskip 1em plus 0.5em minus
  0.4em\relax IEEE, 2008, pp. 435--439.

\bibitem{muller2011chroma}
M.~M{\"u}ller and S.~Ewert, ``Chroma toolbox: Matlab implementations for
  extracting variants of chroma-based audio features,'' in \emph{Proceedings of
  the 12th International Conference on Music Information Retrieval (ISMIR),
  2011. hal-00727791, version 2-22 Oct 2012}.\hskip 1em plus 0.5em minus
  0.4em\relax Citeseer, 2011.

\bibitem{harte2006detecting}
C.~Harte, M.~Sandler, and M.~Gasser, ``Detecting harmonic change in musical
  audio,'' in \emph{Proceedings of the 1st ACM workshop on Audio and music
  computing multimedia}, 2006, pp. 21--26.

\bibitem{mika1999kernel}
S.~Mika, B.~Sch{\"o}lkopf, A.~J. Smola, K.-R. M{\"u}ller, M.~Scholz, and
  G.~R{\"a}tsch, ``Kernel pca and de-noising in feature spaces,'' in
  \emph{Advances in neural information processing systems}, 1999, pp. 536--542.

\bibitem{kominek2008synthesizer}
J.~Kominek, T.~Schultz, and A.~W. Black, ``Synthesizer voice quality of new
  languages calibrated with mean mel cepstral distortion,'' in \emph{Spoken
  Languages Technologies for Under-Resourced Languages}, 2008.

\end{thebibliography}


\end{document}